# IMPROVING THE SCIENCE PROCESS SKILLS OF PHYSICS EDUCATION STUDENTS BY USING GUIDED INQUIRY PRACTICUM

**Albertus Hariwangsa Panuluh**

*Department of Physics Education, Faculty of Teacher Training and Education, Sanata Dharma University, Mrican, Tromol Pos 29, Yogyakarta 55002, INDONESIA*

panuluh@usd.ac.id

**Abstract**

This research investigate that science process skills significantly improve after doing some practicum activities. The research population are fifth semester physics education students and the research sample are fifth semester physics education students who was doing electricity and magnetism experiment C class course. We used two questionnaires, the first one is given to the students after doing three experiment activities and the second one is given after doing six experiment activities. This research is quantitative research using paired sample t test analysis that compared the first questionnaires score and the second questionnaires by using *SPSS* software. The result indicates that the number of practicum activities is able to improve the science process skills significantly.

**Introduction**

The development in the education world especially physics education is very rapid because of the development of technology. Nowadays, the students from elementary until high school can access everything from media for example you can find information about atomic structure lecture from *youtube* or *blog*. But, the media in internet can contain wrong informations. In this case, the role of teachers as educator and companions is very needed.

To begin with, the teacher's skills and mastery of teaching materials should be deepened. Moreover, the scientific approach is used and taught in 2013 curriculum (K13). Not only scientific approach, character building also is introduced so that the students have good knowledge and character. Therefore, the most important and easiest way to reach K13 goals is to improve the science process skill of teacher candidate students in university. Particularly physics teacher candidate.

University plays a role to teach and deepen content knowledge and an appropriate method to teach a concept or teaching material. The offered course in the study program is expected can develope the affective and psychomotor aspect besides the cognitivive aspect.

One physics learning method that is able to combine both aspects is practicum. The practicum method is a method that involves students to be active in an experiments of physics materials that have or will be studied. This practicum method is one of the constructivist methods, i.e the students will find something during the practicum at lab so that students do not memorize but find something. It is hoped that the students will become deeper in understanding the concepts of physics so that they are ready to become teacher in the future.

In practicum, the students are also taught to work in groups, both in assembling, analyzing, discussing and concluding what has been obtained by doing practicum. By using practicum method can also train some characters building such as: rational thinking, cooperation, respect for others, honest, conscientious, disciplined, respect for nature and God. The importance of practicum experiences as a key determinant of pre-service science teachers' emerging inquiry-based science views and practices (Fazio dkk, 2017).

**Theory**

Physics education is a part of sciencle education which have three elements, that are knowledge, process and attitude (Martin, 1991). The knowledge about nature laws and the underlying theory always be emphasized in physics education. Student study physics material for example Newton Law, Relativity, Atomic Theory, etc to improve and apply this material to daily life.

Physics education helps students to know how the physicists work when they did some experiments and made a conclusion from those experiments. This is what we call scientific method. By using scientific method, student is hoped can think rationally and making a conclusion from data that they have collected. Physics education also helps students in developing correct learning attitude, for example honest, discipline, thorough, objective, not doing data manipulation, and teamwork.

Sugianto et al. (2009) said that science process approach is teaching and learning approach which emphasize the study process, activity, and creativity of students when obtain the knowledge, skill, achievement and attidue, and also apply the theory in daily life. Hamalik (1995) said there are six aspects which want to be improved by using science process approach, that is (1) question, (2) hypothesis, (3) investigation, (4) observation, (5) classification, (6) prediction, (7) interpretation and (8) communication.

Suparno (2007) said that, in general experiment methods is teaching method which invite student to do experiment or practicum as verification of the theory. This method also invite students to be active and doing in groups so we can identify which student is active or passive. Hamalik (1983) said that there are some benefits by using experiment or practicum in teaching and learning, (1) exercise to apply the theory that have been learnt, (2) To obtain practical experiences that did not obtain in class, (3) to find another theory.

**Methodology**

This research is quantitative descriptive. The population is the fifth semester of physics education student and the sample is 15 students of electricity and magnetism practicum class C. The instrument used is two questionnaires, the first after doing three practicums and the second after doing six practicums.

This research is a quantitative research with analysis using paired sample t-test. The instrument that used is a Likert-scale questionnaire on aspects of science process skills. The first questionnaire was given to the students after three practicums and the second questionnaire was given after six practicums. The first and second questionnaire scores will be compared using paired sample t-tests with help of the SPSS 22 program to see the significant improvement or not.

**Results and Discussion**

  A.  Science Process Skill Improvement

The mean value from pretest and posttest for six aspects of science process skill is shown in table 1. We can see that the mean value is increase from 76,33 to 82,40 after posttest.

**Table 1.** The science process skill pretest and posttest mean value

|  |  | Mean | N | Std. Deviation | Std. Error Mean |
|---|---|---|---|---|---|
| Pair 1 | pretrampil | 76.33 | 15 | 8.772 | 2.265 |
|  | postrampil | 82.40 | 15 | 9.934 | 2.565 |

**Table 2.** The value of paired sample t-test of science process skill

|  | Paired Differences | | | | | | | |
|---|---|---|---|---|---|---|---|---|
|  |  |  |  | 95% Confidence Interval of the Difference | | | | |
|  | Mean | Std. Deviation | Std. Error Mean | Lower | Upper | t | Df | Sig. (2-tailed) |
| Pair pretrampil I – postrampil | -6.067 | 10.194 | 2.632 | -11.712 | -.421 | -2.305 | 14 | .037 |

Furthermore, with the help of SPSS program will be analyzed whether by doing practicum will be a significant improvement in science process skills. The results of the analysis are presented in Table 2. From Table 2 the analysis using SPSS program, related to the process of science skills experienced a significant improvement. This is indicated by the value of $p = 0,037 < a = 0,05$. So it can be concluded that by doing more experiments will improve students' science process skills.

**B. Science Process Skill Aspects**

There are eight aspects of the science process skills that will be seen in this study: (1) asking, (2) hypothesis, (3) investigation / planning experiment, (4) observation, (5) classification, (6) prediction, (7) interpretation and (8) communication. The mean pretest and posttest values for the eight aspects of the science process skills are presented in Table 3. It can be seen that there is an improvement in mean values for the eight aspects of the science process skills.

For further research, it will be analyzed using paired sample t-test with the help of SPSS 22 to know which aspects of process skill are significantly improved. The results are shown in Table 4. From Table 4 it can be seen that significant increase occurs in two aspects, namely observation aspect ($p = 0,009 < a = 0,05$) and communication aspect ($p = 0,023 < a = 0,05$). While for the six other aspects (ask, hypothesis, investigation, classification, prediction and interpretation) improved but not significant.

Table 3. Pretest and posttest science process skill aspects mean value

|  |  | Mean | N | Std. Deviation | Std. Error Mean |
|---|---|---|---|---|---|
| Pair 1 | Prebertanya | 6.27 | 15 | 1.486 | .384 |
|  | Posbertanya | 6.87 | 15 | 1.552 | .401 |
| Pair 2 | Prehipotesis | 7.27 | 15 | 1.534 | .396 |
|  | Poshipotesis | 7.60 | 15 | 1.352 | .349 |
| Pair 3 | Preinvest | 13.73 | 15 | 2.251 | .581 |
|  | Posinvest | 15.00 | 15 | 2.360 | .609 |
| Pair 4 | Preobserv | 15.20 | 15 | 1.656 | .428 |
|  | Posobserv | 16.40 | 15 | 1.724 | .445 |
| Pair 5 | Preklasif | 8.73 | 15 | .799 | .206 |
|  | Posklasif | 9.07 | 15 | 1.033 | .267 |
| Pair 6 | Prepredik | 6.80 | 15 | 1.082 | .279 |
|  | Pospredik | 7.00 | 15 | 1.000 | .258 |
| Pair 7 | Preinter | 11.00 | 15 | 1.604 | .414 |
|  | Posinter | 12.00 | 15 | 2.204 | .569 |
| Pair 8 | Prekomun | 7.33 | 15 | 1.397 | .361 |
|  | Poskomun | 8.47 | 15 | 1.187 | .307 |

**Table 4.** Paired Sample t-test of science process skills aspect

| | Paired Differences | | | | | t | Df | Sig. (2-tailed) |
|---|---|---|---|---|---|---|---|---|
| | Mean | Std. Deviation | Std. Error Mean | 95% Confidence Interval of the Difference | | | | |
| | | | | Lower | Upper | | | |
| Pair 1 prebertanya - posbertanya | -.600 | 1.882 | .486 | -1.642 | .442 | -1.235 | 14 | .237 |
| Pair 2 prehipotesis - poshipotesis | -.333 | 1.799 | .465 | -1.330 | .663 | -.717 | 14 | .485 |
| Pair 3 preinvest - posinvest | -1.267 | 2.492 | .643 | -2.647 | .113 | -1.969 | 14 | .069 |
| Pair 4 preobserv - posobserv | -1.200 | 1.521 | .393 | -2.042 | -.358 | -3.055 | 14 | .009 |
| Pair 5 preklasif - posklasif | -.333 | 1.113 | .287 | -.950 | .283 | -1.160 | 14 | .265 |
| Pair 6 prepredik - pospredik | -.200 | .676 | .175 | -.574 | .174 | -1.146 | 14 | .271 |
| Pair 7 preinterpre - posinter | -1.000 | 2.360 | .609 | -2.307 | .307 | -1.641 | 14 | .123 |
| Pair 8 prekomun - poskomun | -1.133 | 1.727 | .446 | -2.089 | -.177 | -2.542 | 14 | .023 |

Observation and communication aspects have a significant improve when students do practicum. The more numbers of practicum that students do make them become more proficient in reading tools, especially multimeters because the electric and magnetism practicum almost every practicum using a multimeter. After doing a lot of practicum students become not easily satisfied with the data obtained. They will continue to repeat the data retrieval process. As for the communication aspect, the practicum report written by the student becomes more coherent and clear in making the discussion. As the oral exam of the students is increasingly adept at assembling tools and explaining what was done in the experiment.

## Conclusion

Based on the results of research that has been done the researchers obtained some conclusions as follows.

1. Practicum improves students' science process skills significantly.

2. There are two aspects that experienced a significant increase in aspects of observation and communication.

## Acknowledgment

We want to thank to P4 and LPPM Universitas Sanata Dharma for the support so this research can be done. Also we want to thank to Prof. Dr. Paulus Suparno, SJ, M.S.T for the disscussion.